\documentstyle[12pt]{article}

\newcommand{\EQ}{\begin{equation}}
\newcommand{\EN}{\end{equation}}
\newcommand{\beq}{\begin{equation}}
\newcommand{\eeq}{\end{equation}}

\def\aprge{\buildrel > \over {_{\sim}}}

\def\ref{\hbox{$^{7}$}}

\begin{document}
\hoffset = -0.5truecm
\voffset = -0.5truecm

\topmargin 0pt

\null

\rightline{DOE/ER/40561-136-INT94-13-01}

\vskip 10mm
\begin{center}
{\bf THE SOLAR NEUTRINO PROBLEM:\\
 NEITHER ASTROPHYSICS NOR OSCILLATIONS?}
\footnote{Talk given at the International
Workshop ``Solar neutrino problem: Astrophysics or Oscillations",
Gran \\
\phantom{uiop}Sasso, Italy,
February 28 - March 1, 1994}\\
\vskip 8mm
{Alexei Yu. Smirnov}
\footnote{On leave from Institute for Nuclear Research, Russian Academy of
Sciences, 117312 Moscow, Russia.\\
\phantom{okji}e-mail: smirnov@ictp.trieste.it , smirnov@willy.npl.washington
.edu}\\
\vskip 1mm
{\it Institute for Nuclear Theory, University of Washington, Henderson Hall
HN-12,\\ Seattle, WA 98195,\\
International Centre for Theoretical Physics, I-34100 Trieste, Italy}\\

\vskip 9mm
{\bf
Abstract}
\end{center}
\vskip 1truecm
There is no consistent solar model which
can describe all experimental data on the solar neutrinos.
The problem can be formulated essentially in a model independent way.
The key points are the comparison of the Homestake and the
Kamiokande data as well as the comparison of the GALLEX  and SAGE
results with minimal signal  estimated from the solar luminosity. It is
argued than in such a comparison one should use the Homestake-II data
(only after 1986) with caution. The results of the model independent
analysis  show strong suppression of the beryllium neutrino flux.
The data can be well described by the resonant flavor conversion.
For the ``low flux model" which can
accommodate the Kamiokande signal, a consistent
solution can be found for the neutrino mass  squared difference
$\Delta m^2 = (0.3 - 1.0)\cdot 10^{-5}$ eV$^2$
 and values of mixing angle $\sin^2 2\theta >  5 \cdot 10^{-4}$
(``very small mixing solution").


\newpage
\section{Introduction}

The solar neutrino problem is usually formulated
as  disagreement of the experimental
signals [1 - 4]  with the predictions of the {\it ``reference" standard
solar models} [5 - 9]\footnote{
Some authors however refer to   the
problem as to the impossibility to explain
the data without introducing new neutrino properties.}.
The first GALLEX results (1992) have boosted new attempts to find
non-neutrino physics
solution of the problem, and
 main points are the following.

1. GALLEX and  later SAGE signals exceed the signal from
the standard pp-neutrino flux as well as the minimal signal  estimated from
the luminosity of the Sun.
 The pp-neutrinos compose the bulk of the solar neutrino   flux
and its  prediction is  the most accurate and reliable.
  Some suppression (0.6 - 0.7) of the observed Ga - signal
can be related
to  smaller fluxes of
  high energy neutrinos for which the predictions
are strongly model dependent and not  as reliable as for pp-neutrinos.

2. Kamiokande II+III gives
$$
(signal) = (0.5 - 0.7) \times(SSM),
$$
 i.e. the signal is in agreement with prediction within theoretical
uncertainties estimated as $\sim 40\%$.

3. The energy distribution of the Kamiokande events agrees with undistorted
energy spectrum of the boron neutrinos. However, the experimental
errors  are rather large and the  distortions
 implied by a number of neutrino physics solutions of the
problem can not be excluded.

4. The Homestake signal  after 1986 is rather high: about 3 SNU.
 So that there is no direct contradiction between the Homestake
and the Kamiokande results. The boron neutrino flux extracted
from the Homestake data and the flux measured by Kamiokande
agree within error bars.

5. The difference between the Homestake results before 1986 and after 1986
 may be just a statistical fluctuation.
The results before 1986 show time dependence which can be due to
some unknown systematics.
Comparing the Homestake and the Kamiokande results one should
use the Homestake data  during the time of operation of the Kamiokande
experiment.  In this period the signals from  both experiments
have  no appreciable
time dependence.

6. There are essential uncertainties in the extrapolated
(to solar energies) cross-sections
$\sigma_{17}$, $\sigma_{34}$ . Some latest experimental and theoretical studies
indicate
that the cross sections can be 30 - 40 \% below those used in
 the reference models [10,11].

7.
The collective plasma effects are not taken into
account appropriately. A number of corrections may result in diminishing
of the opacity, and consequently,
of the central temperature of the Sun up to 2 - 3\% [12].

 8. There are another unresolved problems (e.g. $^7Li$-surface concentration
) which may testify for incomplete understanding of
 properties of the Sun (inner convection ?).
This in turn, puts the question mark on the reliability of the solar
flux predictions, although it is unclear what could be the impact of the
solution of the $^7 Li$ problem on the inner structure of
the Sun and on the solar fluxes.

In this paper we will consider  present  status of the
solar neutrino problem.

\section{Data versus predictions}

\noindent
{\bf 2.1 The data}
\vskip 2mm

1). The average  Ar-production rate  during all observation time
(runs 18 - 126) equals after background subtraction \cite{home}
\beq
Q_{Ar} = 2.32 \pm 0.16(stat) \pm 0.21(syst)~~ {\rm SNU}.
\eeq
For the present discussion it is instructive to divide the data into two
parts: the data before pump breaking in 1985 and after resuming the
experiment in 1986. For the sake of
brevity we will call the data before 1986 as data
from Homestake-I and after 1986 as data from Homestake-II. The averaged
signals are:
\beq
\matrix{
Q_{Ar}^{I} = 2. 07 \pm 0.25 ~ SNU ~~(Homestake-I, runs ~~18 - 89)\cr
Q_{Ar}^{II} = 2.76 \pm 0.31~ SNU~~ (Homestake-II, runs ~~90 - 126).
}
\eeq

The latest data do not confirm the anticorrelation with solar activity:
large number of the sunspots in 1990 - 1991 was accompanied by high
counting rate; relatively small signal was observed during quiet 1992.
On the other hand, the data confirm 2 - 3 years period variations of
signal. An impressive increase of the Ar-production rate has been
observed  in the time of solar flare in June 1991. This may testify for
 incomplete understanding of the physics of the Sun or the Homestake
experiment itself.\\

2). The boron neutrino flux measured by
Kamiokande II+III  in the units $\Phi^{SSM}_0$, where
$\Phi^{SSM}_0 \equiv 5.8 \cdot 10^6$ cm$^{-1}$ s$^{-1}$ is
the central value of
flux predicted by the standard solar model
\cite{bahcu} , is
 \cite{kami}
\beq
R_{\nu e}^{II + III} \equiv \frac{\Phi^{exp}}{\Phi^{SSM}_0} =
0.50 \pm 0.04 (stat.) \pm 0.06 (syst.), ~~~ (1
\sigma).
\eeq
The data  agree  with constant neutrino flux. No anticorrelations (or
correlations)
with solar activity
were found. Possible time variations should not exceed  30\%.
The energy distribution of events can be fitted with practically the
same  probabilities by constant and MSW-nonadiabatic suppression factors.

3). The average  Ge-production rate measured by
GALLEX-I+II is  \cite{gall}
\beq
Q_{Ge}^{I+II} = 79 \pm 10(stat) \pm 7(syst)~~(1\sigma~ {\rm prelim.)}.
\eeq
 The  combined error is $\sim$ 12 SNU.
The data are $\sim 3\sigma$ below the expected value 125 - 130 SNU.

4). Statistical analysis of the Ge-production rate in 15 runs
of the experiment SAGE gives \cite{sage}
\beq
Q_{Ge} = 74 \pm 19(stat) \pm 10(syst)~ SNU ~~(1\sigma ~~prelim.).
\eeq

\vskip 3mm
\noindent
{\bf 2.2 Comparison with model predictions}
\vskip 2mm

A comparison of experimental data with
 model predictions is given in Fig.1. There are several immediate
observations.

All the experiments have  detected signals  which are lower or
much lower
than the predictions of the ``reference" standard solar models [15 - 19].

New measurements of nuclear cross-sections as well as plasma effects
revision probably will result in  appreciable reduction of the boron
neutrino flux. However, even the
``low flux models"  [13, 14]
which could accommodate the Kamiokande  result
predict the argon production rate $Q_{Ar} \sim (4.2 - 4.5$) SNU, whereas
the experimental value of $Q_{Ar}$ averaged over all observation time is
$\approx 8\sigma$ smaller.
For  Ge-production rate these models give about 110 SNU,
new GALLEX result is $2.5\sigma$ lower (fig.1).

Suppression of signals in different experiments is different. The
Homestake signal  is suppressed  stronger than  the
Kamiokande one:  the double ratio, \\
$(data/model)_{Hom}/(data/model)_{Kam}$, is
practically the same for all models:
\beq
\frac{R_{Ar}}{R_{\nu e}} = \left\{ \matrix{ 0.58 \pm 0.12, ~~[6]\cr
                                       0.57 \pm 0.12, ~~  [7] \cr
                                       0.58 \pm 0.11, ~~  [14] \cr
} \right .  .
\eeq
This statement can be relaxed if one takes  the Cl- Ar data only
for a period of the operation of Kamiokande:
$\frac{R_{Ar}}{R_{\nu e}} = 0.69 \pm 0.22~$.

The difference in $R_i$ testifies for the energy dependence of the
suppression effects.

\section{Homestake versus Kamiokande}

\noindent
{\bf 3.1 Model independent comparison. Contradiction?}
\vskip 2mm
One can perform a direct test of  consistency of the
Cl - Ar and Kamiokande results in the model independent way \cite{bara}.
Suppose that

1). there is no distortion of the  energy spectrum of boron neutrinos and

2).
the  Kamiokande signal is due to the electron neutrino  scattering only.

 In this case  Kamiokande measures immediately
the flux of the electron neutrinos from boron decay:
\beq
\Phi_B = (2.9 \pm 0.42) \cdot 10^6 ~~cm^{-2} s^{-1}.
\eeq
 This flux gives the contribution
to  Ar production rate:
\beq
Q_{Ar}^{B} = 3.1 \pm 0.45 ~~{\rm SNU}
\eeq
which exceeds the  total  measured
rate by $\sim 2\sigma$:
$Q_{Ar}^B > Q_{Ar}^{total} .$
The inclusion of the contributions from Be- and
other neutrinos strengthens the disagreement.

Main objection is that one should use the data only from Homestake-II
when compare with Kamiokande.
The result (8) agrees with Homestake-II signal (2)
within $1\sigma$ if
the contributions from all other neutrinos are strongly suppressed.

\vskip 3mm
\noindent
{\bf 3.2 Boron neutrino flux from Homestake and Kamiokande}
\vskip 2mm

One can confront the Homestake and Kamiokande data  comparing the
boron neutrino fluxes measured by Kamiokande and {\it extracted} from the
Homestake experiment. It was claimed in  [1, 14] that the
Kamiokande flux and
the flux extracted from Homestake-II are in agreement.
However this statement is the result of {\it selection} of the data.
Indeed, in [1]

1). the data from Kamiokande-II where used only,

2). the threshold $E^{th} = 9.3 $ MeV was chosen,

3). it was suggested that the contribution to $Q_{Ar}$ from the boron
neutrinos is $r_B  \equiv Q_{Ar}^B/Q_{Ar} = 0.77$ of all signal.
 This  number corresponds to
the Bahcall-Ulrich model. However for the models with low flux of
 boron neutrinos the contribution of boron neutrinos is typically
smaller: $r_B = 0.69 - 0.70$.

Using the data from both the Kamiokande-II and the Kamiokande-III, the
threshold $E^{th} = 7.5$ MeV and the boron  contribution $r_B = 0.7$, one
gets the picture shown in fig.2. The flux extracted from the
Homestake-II is systematically lower than that measured by Kamiokande.
The average value of flux is
\beq
\Phi_B(H-II) = (1.8 \pm 0.2)\cdot 10^{6}~~ cm^{-2} s^{-1}
\eeq
i.e.
about $2.5\sigma$  lower than
the Kamiokande flux (7).
This situation can be described also in ``boron - beryllium" neutrino plot
(see sect.5.1).

\vskip 2mm
\noindent
{\bf 3.3 Statistical fluctuation?}
\vskip 2mm

 In principle, it is correct  to use only the Homestake-II data
 when comparing with Kamiokande. However, let us consider the
Homestake-II results more carefully.
In fig.3a  is shown the distribution of number of runs $n$  with a given
Ar-production rate $N_{Ar}$: $n = n(N_{Ar})$ for Homestake-II.
Fig. 3b and 3c. show the distribution of  the Homestake-I
runs  as well as the result
of Monte Carlo simulation for average production rate corresponding to
$Q_{Ar} = 2.3$ SNU.  As follows from these figures the shape of  the
Homestake-II distribution disagrees with both Monte Carlo and with the
Homestake-I results.
Note that the latter is well described by
 simulations.
The distribution of Homestake-II runs has two components:
one component is in a good agreement with simulation for $Q_{Ar} = 2.3$
SNU. Another component is a thin peak in the interval $N_{Ar} = 0.7 - 0.9$
at/day, i.e. there is an excess of runs with high counting rate.
14 runs (among 34) where found in the indicated interval, whereas the
simulation gives 4 runs only.
Also the shape of the Homestake-II distribution can not be reproduced
by Monte Carlo simulation with average $Q_{Ar} = 3$ SNU.

For comparison we show the corresponding distribution for the
GALLEX experiment (fig. 3d),
where the number of runs (30) is about that in Homestake-II.

What could be the interpretation of the excess observed by Homestake-II?
The peak in the  distribution can be  a statistical fluctuation.
In this case one expects in future
that average value after 1986 will approach present average
value for all runs.  The peak could be a result of some systematics
(background ?). After subtraction of the peak one gets the result which
agrees
with average value 2.3 SNU.

Using the Homestake-II result one should explain the shape of the distribution
as well as the difference of the Homestake-I and Homestake-II signals, i.e.
the change of the signal with time.
\newpage

\section{Splitting of cycle. Cross sections. Luminosity normalization}

The important relations and dependencies concerning the solar neutrino
fluxes can be obtained immediately from some
general astrophysical notions without modeling.

\vskip 3mm
\noindent
{\bf 4.1 Splitting of the cycle}
\vskip 2mm

The pp-cycle is splited: there are several branches of the reactions
(fig.4 ). It is this branching that results in the uncertainties of the
predictions of the neutrino fluxes.
No branching - no (or almost no) problem. Indeed, without splitting
the fluxes  of the neutrinos produced in the first, say pp-,
and the subsequent, say B-decay, reactions are equal and
 coincide with number of chains per second:
$\Phi_{pp}$ = $\Phi_{B}$ = N. (Strictly speaking this implies also the nuclear
reactions equilibrium, i.e. that there is enought time for the termination
of chains).
Moreover, if the Sun is in thermal equilibrium,  the number
of chains is fixed by the total solar luminosity.
It is the branching that  depends on the solar conditions
as well as on  the nuclear cross-sections.

{\it $^3 He$ -branching}. There are two main possibilities for
$^3 He$
: to interact with another nuclei $^3 He$ which means that in
this chain the second pp-neutrino is produced or to interact with $^4 He$,
producing  $^7 Be$,  and consequently beryllium or boron neutrinos.
The ``branching ratio" $r$ is determined by corresponding
cross-sections and concentrations ($n_3$, $n_4$):
\beq
\frac{r}{1 - r} = \frac{<\sigma_{34}v>}{<\sigma_{33}v>}\cdot
\frac{n_4}{n_3}.
\eeq
Since in both reactions the electric charges of nuclei
are the same,
the dependence of branching on the temperature of the Sun is rather
weak.
The most important dependence  of the branching is that on the
cross-sections (astrophysical factors).
The concentration of the $^3 He$ itself depends on the
cross-section $\sigma_{33}$. Main channel
of the $^3 He$ disappearance is the
$^3He + ^3He$ -
reaction; its probability is proportional to
$W \propto n_3^2 \sigma_{33}$ , therefore the
equilibrium concentration of $^3 He$ equals
$n_3 \propto 1/\sqrt{\sigma_{33}}$.
Adding the  effect of  another channel, $^3He + ^4He$, gives
$n_3 \propto (\sigma_{33})^{-0.56}$.
Substituting this relation in (10) one finds the dependence of
branching ratio on the cross-sections (astrophysical factors)
\cite{cast} :
\beq
r \propto \frac{
\sigma_{34}}{\sqrt{\sigma_{33}}}
\eeq

{\it $^7 Be$ - branching}. $^7 Be$ can capture
the electron, emitting the
beryllium neutrino, or can interact with proton, producing the boron - 8 which
in turn decays with emission of the boron neutrino. The branching ratio,
$r'$,
\beq
\frac{r'}{1 - r'} = \frac{<\sigma_{17} v>}{W_e}
\cdot \frac{n_1}{n_e}
\eeq
strongly depends on the temperature due to  the Coulomb barrier
for the $p- ^7 Be$-reaction. Evidently it is proportional to the
 cross section $\sigma_{17}$. In (12) $n_1$ and $n_e$ are the concentrations
 of the protons and the electrons.

Using the parameters $r$, $r'$ one can find the following relations
between the number of chains and the fluxes of different neutrinos:
\beq
\Phi_{pp} = \frac{N}{2}(2 - r)~, ~~
\Phi_{Be} = \frac{N}{2}r(2 - r')~, ~~
\Phi_{B} = \frac{N}{2}rr',
\eeq
(compare with the toy case).

\vskip 3mm
\noindent
{\bf 4.2 Solar luminosity and the normalization of the neutrino
flux}
\vskip 2mm

The chain of the nuclear reactions
 results in hydrogen burning, production of the $^4He$
\beq
4p + 2 e^- \rightarrow ^4 He  + 2\nu_e + Q,
\eeq
and
energy release $Q = M(^4 He) - 4 M_p + 2 M_e \approx 26.7$ MeV.
It is supposed that the Sun is in thermal equilibrium, i.e.
total luminosity of the Sun  equals the nuclear energy release $Q_N$.
If the energy release
is approximately constant then one can  write the equality
\beq
L_{\odot} = Q_N - L_{\nu},
\eeq
here $L_{\nu}$ is the neutrino luminosity.
 The total energy release and
the neutrino luminosity can be expressed in terms of the neutrino fluxes
$\Phi_i$ , i = pp, Be, B, ... as
\beq
Q_N = \frac{Q}{2}
\sum_{i} \Phi_i ~, ~~~~
L_{\nu} = \sum_{i} E_i \Phi_i ~,
\eeq
where in the first equality we have taken into account that in
each chain of reactions two neutrinos are emitted.
 In the second equality
$E_i$ is the average energy of the neutrino from the i-reaction.
Substituting (16) in (15) one gets
the desired  luminosity normalization condition for the
neutrino fluxes:
\beq
 \sum_{i} \left(\frac{Q}{2} -  E_i \right) \Phi_i = L_{\odot}.
\eeq
Note that according to the SSM the pp-neutrinos compose practically 93\% of the
sum in (17), Be-neutrinos give only 7\%, the contributions from other neutrinos
are negligible.
 Therefore the luminosity condition is sensitive
 mainly to the pp-, and the beryllium neutrinos. Neglecting also the
contribution from the $\nu_{Be}$ one gets the estimation of the
pp-neutrino flux:
\beq
\Phi_{pp} \approx \frac{2 L_{\odot}}{Q - 2 E_{pp}}.
\eeq

\vskip 3mm
\noindent
{\bf 4.3 Time variations of the energy release?}
\vskip 2mm

One remark is in order.
Using the equality (15) one should keep in mind
the difference in time.
Present (electromagnetic) luminosity of the Sun $L_{\odot}$ is
determined by  energy release about $t_d \sim 10^6 - 10^7$ years ago, and
more precisely,
the condition of thermal equilibrium should be written as
\beq
L_{\odot}(t) = Q_N(t - t_d) - L_{\nu}(t - t_d).
\eeq

The variations of the  energy release on the
time scales $t \ll t_d$  are  averaged out and
luminosity gives an information  about the averaged energy
release and the average neutrino fluxes.
 In the Standard Solar Models there is no appreciable changes of
$Q_{N}$ during $10^7$ years. However, short term variations
of
$Q_N$ ($10^5$ years, 22 years, 2 years, months, hours?) related to some
instabilities in the core of the Sun are not excluded.

\section{Analysis of  all data}

{\bf 5.1 ``$\nu_B - \nu_{Be}$"-plot}
\vskip 2mm

It is convenient to analyze the data using the
$\nu_B - \nu_{Be}$-plot
-- the plot of the boron and beryllium neutrino
fluxes \cite{cast}, \cite{blud} (fig.5).
Let us measure the neutrino fluxes, $\Phi_i$ in the units of fluxes
$\Phi_i^0$, predicted by a certain reference SSM:
$$
\phi_i \equiv \frac{\Phi_i}{\Phi_i^0}.
$$
For definiteness we take for $\Phi_i^0$ the central values of
Bahcall-Ulrich model. In terms of $\phi_i$ the signals in different experiments
can be written as:
\beq
R_{\nu e} = \phi_B,
\eeq
\beq
Q_{Ar} = 6.1 \phi_B + 1.1 \phi_{Be} + Q_{Ar}^{other},
\eeq
\beq
Q_{Ge} = 14 \phi_B + 34 \phi_{Be} + 71 \phi_{pp} + Q_{Ge}^{other}.
\eeq
In (20) it was suggested that there is no distortion of the boron neutrino
spectrum and the signal is stipulated by the electron neutrino only.
In (21,22) the numerical coefficients  (in SNU's)
 correspond to chosen reference
model. For fixed reference model the coefficients are determined by
the cross-sections of the neutrino interactions in detectors.
$Q_{Ar}^{other}$ and $Q_{Ge}^{other}$ are the contributions from other fluxes
which are typically smaller than the explicitly indicated
contributions.

The flux of the pp-neutrinos can be extracted from the $L_{\odot}$
normalization
condition. Neglecting all the contributions in (17) apart from pp- and
Be- contributions one  gets
\beq
\phi_{pp} + k \phi_{Be} = 1 + k,
\eeq
where
$$
k \equiv \frac{Q - 2 E_{Be}}{Q - 2 \bar{E}_{pp}} \cdot
\frac{\Phi_{Be}^0}{\Phi_{pp}^0},
$$
and $k \approx 0.075$ in SSM [5]. Substituting (23) in (22) one finds:
\beq
Q_{Ge} - Q_{Ge}^{other} - 71 (1 + k) = (34 - 71 k) \phi_{Be} +
14 \phi_B.
\eeq
 The experimental data  on $Q_{Ar}$, $Q_{Ge}$ and $R_{\nu e}$
and the estimations  of $Q^{other}$  give according to
(20, 21, 24) the allowed regions (strips) for each experiment  shown
on the ``boron - beryllium" plot.

\vskip 3mm
\noindent
{\bf 5.2 Confronting all data}
\vskip 2mm

In fig.5 are shown the regions allowed at $1\sigma$ level
by the Homestake, Kamiokande and GALLEX
experiments. We have suggested that
$Q^{other} =  Q^0/2$. Zero values of $Q^{other}$ slightly relax
the bounds (see below).
As follows from fig.5  the Homestake and Kamiokande results imply
strong  suppression of the $\nu_{Be}$- flux. There is no intersection
of corresponding $1\sigma$-allowed regions in the
$\nu_B - \nu_{Be}$-plot.
The intersection of $2\sigma$-allowed
regions appears for values of $\Phi(\nu_{Be})$ being
2.5 times smaller than the  predictions of the reference models.

The disagreement between the Kamiokande and Homestake relaxes (but does not
disappear) if one takes the Homestake-II results.
In this case the intersection
of $1\sigma$-allowed regions appears for $\Phi(\nu_{Be}) < 10^9 cm^{-2}
s^{-1}$, i.e.  for 5 times smaller  flux than reference models predict.
Moreover, as we have mentioned  the data after 1986 show some statistical
inconsistency.

New GALLEX-I+II results
have  small errors which allow to make some
important conclusions. GALLEX signal is just slightly higher than the
signal expected from the pp-neutrinos. Therefore if there is no
conversion of pp-neutrinos for which one has rather accurate predictions, the
GALLEX data testify for strong suppression of the contributions from
all other components of the neutrino flux and first of all from the beryllium
neutrinos.

The intersection of $1\sigma$
regions  allowed by GALLEX and Kamiokande on
``$\nu_B - \nu_{Be}$"-plot gives the same  (factor of 5)
suppression of $\Phi(\nu_{Be})$ as  Homestake and Kamiokande give.

\section{Neither astrophysics nor oscillations?}

\noindent
{\bf 6.1
 Astrophysics}
\vskip 2mm

The astrophysical solutions fit the conditions formulated in sect. 3.1
   and consequently,
meet the problems discussed above.
A number of  modifications of  solar models
were suggested which result in decrease of the central temperature of the Sun,
$T_c$. However,  $T_c$ decrease suppresses the boron
neutrino flux stronger than the beryllium neutrino flux,
and consequently, the double ratio in (6)
becomes even smaller.
Essentially for this reason a combined fit of all the data
for arbitrary astrophysical parameters is rather bad -
any astrophysical solution  is  excluded at 98\% C.L. \cite{blud}.

\vskip 3mm
\noindent
{\bf 6.2 Nuclear physics solution}
\vskip 2mm

Using (10) one can estimate the dependence of the fluxes on the cross-
sections:
\beq
\frac{\Delta \Phi_{Be}}{\Phi_{Be}} \sim (1 - r) \frac{\Delta \sigma_{34}}
{\sigma_{34}}.
\eeq
Precise study gives the coefficient 0.81 [5]. Boron neutrino flux
has similar dependence on $\sigma_{34}$. Evidently $\Delta \Phi_B /
\Phi_B \sim \Delta \sigma_{17}/\sigma_{17}$.
 A decrease of the  astrophysical factor
$S_{17}$ allow to suppress the boron neutrino flux without
changes of the Be-neutrino flux as well as  the solar model.
However,  present data testify for strong suppression of $\phi_{Be}$.
 Certainly,
30 - 40 \% reduction of the  $S_{34}$ is not
enough to solve the problem.  According to (fig.5)
the desired suppression of Be-branch of the pp-cycle implies strong
(factor  3 - 5) suppression of the astrophysical factor $S_{34}$ or even more
strong (10 times) increase (see (11)) of
$S_{33}$ \cite{cast}.
An appreciable increase of $S_{33}$ could be related to the existence of
the hypothetical $^3$He - $^3$He resonance.
Obviously,
strong suppression of the Be-branch gives also strong reduction
of $\nu_B$-flux.
 The  ``astrophysical" suppression of the $Be$-flux may
imply an essential modification of solar models,
since Be-neutrinos are related (according to (17))
to $\approx  7 \%$ of the Sun luminosity
fixed with $0.2\%$ accuracy. The plasma effects are basically  reduced
to  diminishing of the central solar temperature  which in turn
results in more strong suppression of $\nu_B$-flux than $\nu_{Be}$-flux,
i.e. does not solve the problem.

The increase of the cross-section of the pp-reaction works essentially
as the decrease of the central temperature \cite{blud}.

Formally one could suppress the beryllium line by
strong diminishing $\sigma_{34}$ (or increasing $\sigma_{33}$),
the corresponding decrease of the boron flux could be compensated by
increase of $\sigma_{17}$ or/and the temperature.

\vskip 3mm
{\bf 6.3 Large, small, very small?}
\vskip 2mm

All the data obtained so far can be easily
reconciled with predictions of the reference standard solar models
by the resonant flavor conversion (MSW-effect) \cite{msw}  $\nu_e \rightarrow
\nu_{\mu} (\nu_{\tau})$.
The best description of the data can be obtained for small mixing
angles when the suppression pit is rather thin, so that the
pp-neutrinos are outside the pit, the Be-neutrinos are at the bottom and
the boron neutrinos are on the non-adiabatic edge (see fig. 6).
Note that in this case  the conditions of sect. 3.1 are broken:
the
Homestake and Kamiokande data can be reconciled
due to more strong suppression of low energy part of the boron neutrino
spectrum which does not contribute to the Kamiokande signal
and due to  an additional contribution
to Kamiokande from the scattering of the converted $\nu_{\mu}$ ($\nu_{\tau}$)
via neutral currents.
For the reference models  [6] the data pick up the
 region of parameters (see fig. 7):
\beq
\Delta m^2 = (0.5 -1.2) \cdot 10^{-5}~ {\rm eV}^2,~~~
\sin^2 2\theta = (0.3 - 1.0) \cdot 10^{-2}.
\eeq
 Also the region of large mixing solution is not excluded:
\beq
\Delta m^2 = (1 - 3) \cdot 10^{-5}~ {\rm eV}^2,~~~
\sin^2 2\theta = (0.65 - 0.85).
\eeq

Although the solar neutrino problem can be formulated in  a model independent
way the implications of the results,  and in particular, the appropriate
regions of the neutrino parameters depend on the predicted fluxes.
 The boron neutrino flux being strongly involved in this determination
has rather large uncertainties. In the model [6] they estimated to be
at the level 40\%.  The changes of the cross sections and plasma
effect revision may reduce the predicted boron neutrino flux
 to that measured by  Kamiokande.
 From this one can conclude that

 1). the uncertanties  of the $\nu_B$-flux can
not be considered as just the statistical ones,

2). probably, these uncertainties  will
not be essentially diminished in near future
(at least to the moment  when
new  solar
neutrino experiments will start to work),

3). one should try to solve the problem without referring to the original
(theoretical) value of the
boron neutrino flux. This flux can be considered as {\it free parameter}
which should be determined from the experiments.

Let us  discuss
 in this connection the solution of the problem
in the context of the  ``low flux models" which predict
the boron neutrino flux at the level of that measured by Kamiokande.
In other words let us  suppose that $R_{\nu e} \rightarrow 1$.
 Strong suppression of the beryllium flux and  weak suppression of
the pp-  as well as  the
boron neutrino  fluxes  can be obtained for very small mixing angles when
the high energy part of the boron neutrino spectrum is at the top,
whereas the beryllium neutrinos are at the bottom of the
nonadiabatic edge (see fig. 6).

For small mixing angles ( down to $\sin^2 2\theta \sim 3\cdot 10^{-4}$)
the  nonadiabatic edge  can be well described
by the Landau-Zener formula [19]:
\begin{equation}
P_B = P_{LZ} \equiv exp\left(-\frac{E_{na}}{E}\right),
\end{equation}
where
\beq
E_{na}   = \frac{\Delta m^2 l_n\sin^22\theta}{\cos2\theta}
\eeq
and $l_n \equiv |\frac{d}{dx}\ln n_e|$. Taking into account the effect
of the neutral currents one has
\beq
R_{\nu e} \approx P_B + \frac{1}{7}(1 - P_B).
\eeq
Using (28, 29) one finds the desired value of mixing angle:
\beq
\sin^22\theta   \approx \frac{E_B}{\Delta m^2 l_n} \ln P_B,
\eeq
where $E_B \sim 10$ MeV is the average energy of the detected part of the boron
neutrino spectrum.
The suppression of the beryllium neutrinos can be found from
\beq
P_{Be} = \left( P_{B} \right) ^{E_B/E_{Be}}.
\eeq

According to (31),  with increase of $P_B$
the allowed region of the neutrino parameters  is shifted  to small mixings
by factor $\ln P_B / \ln P_B^0$. For
$P_B = 0.8$ or 0.9 one gets the factors 4 and 9 correspondingly and
\beq
\sin^2 2\theta = (0.5 - 1.5) \cdot 10^{-3}.
\eeq
The values of the mass difference,
$\Delta m^2 \sim (0.3 - 1) \cdot 10^{-5}$  eV$^2$,  are fixed
essentially by the condition $E_{pp}^{max} < E_a < E_{Be}$, i.e. that
the adiabatic edge ($E_a$)
of the pit is between the highest energy of the pp-spectrum
and the energy of the beryllium line.

The ``very small mixing" solution
is characterized by weak distortion of  high energy
part of the boron neutrino spectrum  and by small effect of the
neutral currents
in this energy region. Indeed, according to (28)
the change of the suppression factor,
$\Delta P_B \equiv P_B(E_2) - P_B(E_1)$,
in the energy interval $E_1 - E_2$
is
\beq
\Delta P_B =  P_B(E_2) - \left(P_B(E_2)\right)^{E_2/E_1}.
\eeq
 Suppose $E_2/E_1 = 2$, then using (34) and (30) one gets
$\Delta R_{\nu e} = 0.07$
for $P_{B} = 0.9$,
whereas
$\Delta R_{\nu e}  = 0.21$ for  reference value $P_{B} = 0.5$.
The contribution of the neutral currents to the $\nu e$ scattering
is $\approx 6$ times smaller than in case of ``reference" models.
 Therefore, measuring the distortion of
 the energy spectrum and the effects of
neutral currents  in future experiments
one can find the values of neutrino
parameters and the original flux of boron neutrinos (for details see [20]).

Note that the region of very small angles (33)  may be natural for mixing
of the first and third generations so that the solar neutrino deficit
could be  explained by the
$\nu_e \rightarrow \nu_{\tau}$  conversion.
Such a scenario can be realized in the supersymmetric $SO(10)$  with
unique scale of symmetry violation.  Although mixing (33)
can be described without fine tuning by formula
\beq
\theta_{e \mu} = \left |\sqrt \frac{m_e}{m_{\mu}} - e^{i\phi}
\theta_{\nu} \right |,
\eeq
which corresponds to
the
$\nu_e - \nu_{\mu}$  mixing,
here $m_e$ and  $m_{\mu}$ are the  masses of the electron and muon,
$\phi$ is a phase and $\theta_{\nu}$ is the angle related to
 diagonalization of the neutrino mass matrix.
The relation (35) between the angles and the masses
is similar to the relation in  quark sector  which
follows naturally from the Fritzsch ansatz for mass matrices.
Such a possibility can be realized in terms of the
see-saw mechanism of the neutrino mass generation.

\vskip 3mm
\noindent
{\bf  6.4 ``Detection solution"}
\vskip 2mm

It is not excluded that the problem has the
``detection" solution, i.e. that some
experimental results are interpreted incorrectly.

The
GALLEX results are rather stable and convincing. It is difficult to
expect  appreciable changes of numbers.
The calibration experiment may result
probably in increase of the  measured neutrino flux. The SAGE experiment
confirms the GALLEX results.

Kamiokande results are stable  and the experiment had been (at least partly)
calibrated.

Homestake experiment   shows the strongest suppression of signal.
There is no calibration. Some  features of the data
have small statistical probability, e.g.,
very low signal during 1978 (five runs
with near to zero counting rate),
 high signal during 1986 - 1992,
the peak in the distribution ``number of runs with a given counting rate"
in the region $N_{Ar} \sim 0.7 - 0.8$ at/day.
There is no explanation of the increase of signal during some solar
flares.

What kind of changes in the existing experimental
data could make the non-neutrino physics solution  preferable?

(i). The signal from Cl - experiment at the level
$Q_{Ar} \aprge 4$ SNU
certainly changes the status of the problem.
Such a situation could be  explained by lower central solar temperature
(e.g. as the consequence of the
plasma effects revision) or/and  by the astrophysical factors
$S_{17}$ and $S_{34}$ being (30 - 40)\% smaller. In this case for the Gallium
experiments one expects
$Q_{Ge} \aprge 100 $ SNU, i.e. $2\sigma$ higher than GALLEX result.

(ii). Suppose that the $\nu e$-scattering experiment gives the  boron
neutrino flux suppressed by
factor of 5  (instead of 2) in comparison with reference model flux.
If then the Be-neutrino flux is diminished by 40\%,  one gets the  Ar-
production rate $Q_{Ar} \sim 2.3$ SNU in agreement with Homestake result.
For the Ga-experiment  again a large effect is predicted:
$Q_{Ge} \aprge 100 $ SNU.
The indicated situation can be reproduced by both the decrease
of the central temperature and the cross-sections (especially,
$S_{17}$).

Let us underline in this connection the importance of  present  GALLEX
result  as well as  further diminishing of the experimental errors.

\section{Conclusion}

\hskip 6mm 1. There is no consistent solar model which can explain all
existing experimental data on the solar neutrinos.

2.
Boron neutrino flux measured by Kamiokande II+III
gives the argon production rate which {\it exceeds} total signal measured
by the Homestake experiment. (It is supposed that the Kamiokande
signal is due to $\nu_e e$-scattering). Including the
effects of neutrinos from other reactions  strengthens
the disagreement.

3.  The
disagreement between the Kamiokande and Homestake results
relaxes (but does not
disappear) if one takes the Homestake results after 1986, i.e.
during  the time
of operation of Kamiokande II+III. The boron neutrino flux extracted
from the Homestake-II data is about $2.5\sigma$ below the
Kamiokande flux. However one should use the Homestake data alone with
caution. The Homestake-II signal is appreciably higher than the
Homestake-I  signal.
Moreover,
the {\it shape} of the distribution: number of runs
with a given Ar-production rate, $n = n(N_{Ar})$, after 1986 does not agree
with Monte-Carlo simulation which  describes rather well  the
distribution of   Homestake-I runs. After 1986
the distribution  has a
 thin
peak in the interval $N_{Ar} = 0.7 - 0.9$ at/day.
If the excess
is the statistical fluctuation,   one expects in future
 the convergence of the Homestake-II results to the
average value for  all  runs.
The peak could be a result of some systematics (background
?); its removing gives the average after 1986 in
agreement with the average during all observation time.

The
Homestake and  Kamiokande results imply strong
suppression of the $\nu_{Be}$- flux.
New GALLEX-I+II results with smaller errors give  the important bounds.
They (as well as SAGE results) are at the level of minimal
signal  which follows from the luminosity normalization condition. The
signal practically coincides with  sum of signals  induced
by unsuppressed pp-neutrino flux (according to luminosity normalization)
and by  boron neutrino flux as measured by Kamiokande.

6.
Present data testify for strong suppression of the beryllium neutrino flux.
Certainly,
30 - 40 \% reduction of the astrophysical factor $S_{34}$ is not
enough to solve the problem.
The desired suppression of Be-branch of the pp-cycle implies
strong increase of $S_{33}$ ($^3$He - $^3$He resonance ?).
Plasma effects are basically  reduced
to the decrease  of the central solar temperature  which in turn
results in more strong suppression of $\nu_B$-flux than $\nu_{Be}$-flux,
i.e. does not solve the problem.

7.
New measurements of nuclear cross-sections as well as plasma effects
revision will result probably in essential reduction of the boron
neutrino flux. However, even
``Minimal flux models" which could accommodate the Kamiokande flux
predict the argon production rate $Q_{Ar} \sim (4.2 - 4.5$) SNU, whereas
experimental value of $Q_{Ar}$ averaged over all observation time is
$\approx 8\sigma$ smaller.
For germanium production rate these models give about 110 SNU,
new GALLEX result is $2.5\sigma$ lower.

8.
The situation when the  fluxes of the pp-neutrinos and the boron neutrinos are
unsuppressed   (or weakly suppressed),  whereas the beryllium neutrino flux
is strongly suppressed can be easily reproduced by resonant flavor
conversion (MSW). The solution  corresponds to   thin suppression pit
(as function of energy)
which is realized at $\Delta m^2 \sim (0.3 - 1)\cdot 10^{-5}$ eV$^2$ and
small values of  mixing angle $\sin^2 2\theta = (0.8 - 1.5)\cdot 10^{-3}$.
In this case for high energy part of $\nu_B$
spectrum one predicts a weak distortion and small effect of neutral currents.

 9.
It is not excluded that the problem may have a ``detection solution", i.e.
that interpretation of the results of some (one ?) experiments is
incorrect.

\vskip 5mm
\noindent
{\Large \bf Acknowledgments}

I  am grateful to V.S. Berezinsky, S.Degl'Innocenti  and P.I.Krastev
 for valuable discussions.  I thank the Institute for Nuclear Theory
at the University of Washington for hospitality. This work was supported by
the U.S. Department of Energy under Grant DE-FG06-90ER40561.

\newpage

\vskip 1cm
{\Large \bf Figure Captions
} \vskip 3mm

\noindent
Fig. 1.  Comparison of the observed signals (hatched regions)
with predictions of different solar models:
1, 2 - Bahcall-Pinsonneault (with and without diffusion) \cite{bahcp},
 3 - Castellani et al. \cite{cast},
4 - Turck-Chieze - Lopez \cite{turc},
 5 - Bertomieu et al \cite{bert},
6 -
Schramm and Shi \cite{schr},  7 - Dar and Shaviv \cite{dars}.

\vskip 3mm
\noindent
 Fig. 2. A comparison of the boron neutrino flux measured by Kamiokande-II+III
(solid)
with that  extracted from Homestake results (dashed).

\vskip 3mm
\noindent
 Fig. 3. The number of runs of the Homestake experiment with a given
Ar-production rate.
a). Homestake-II (runs 90 - 126) ; b).  the Homestake-I runs;
c). Monte-Carlo simulation (from [1]). In fig.3d
 we show for comparison the corresponding
distribution of runs for GALLEX experiment  (30 runs).

\vskip 3mm
\noindent
Fig. 4. Branching of the pp-nuclear reaction chain.

\vskip 3mm
\noindent
Fig. 5.
$\nu_B - \nu_{Be}$-plot
. The lines restrict  1$\sigma$-regions allowed
by the Homestake, GALLEX and Kamiokande experiments. Figures at the curves
are the Ge-production rate in SNU's. Dotted lines show the 1$\sigma$
region allowed by Homestake measurements after 1986 (Homestake-II).

\vskip 3mm
\noindent
Fig. 6.
 The suppression factor due to the MSW-effect as function of the
neutrino energy for different values of
$\sin^2 2\theta$ (figures at the curves).
Also shown is the
energy spectrum of
solar neutrinos (hatched).

\vskip 3mm
\noindent
Fig. 7.
The $\Delta m^2 - \sin^2 2\theta$ regions of the solutions
of the solar neutrino problem
for different values of the original boron neutrino flux:
$\Phi_B = r \Phi_B^{SSM}$, where $\Phi_B^{SSM}$ is from [6].

\end{document}